\documentclass[twocolumn,amssymb,nobibnotes,aps,prl,superscriptaddress]{revtex4-2}
\usepackage{graphicx}
\usepackage{textcomp}
\usepackage{amsmath}
\usepackage{commath}
\usepackage{color}
\usepackage[dvipsnames]{xcolor}
\usepackage{textcomp}

\newcommand{\compatibility}{\mathbf{C}}

\newcommand{\Disp}{\mathbf{U}}

\newcommand{\Elong}{\mathbf{E}}
\newcommand{\sites}{\mathbf{r}}

\newcommand{\xiO}{\xi_1}

\newcommand{\latVecO}{\mathbf{\ell}_1}
\newcommand{\latVecT}{\mathbf{\ell}_2}
\newcommand{\site}{\mathbf{r}}
\newcommand{\siteKag}{\bar{\mathbf{r}}}
\newcommand{\bond}{\mathbf{b}}
\newcommand{\bondKag}{\bar{\mathbf{b}}}

\newcommand{\paramDeform}{\gamma}
\newcommand{\paramRot}{\bar{\gamma}}
\newcommand{\winding}{\nu}
\newcommand{\DK}{BK}
\newcommand{\SK}{SK}
\newcommand{\keepOrDiscard}{\textcolor{black}}

\newcommand{\updated}{\textcolor{black}}

\begin{document}
	\title{Cell augmentation framework for topological lattices}
	\author{Mohammad Charara}
	\author{Stefano Gonella}
	\email{sgonella@umn.edu}
	\affiliation{Department of Civil, Environmental, and Geo- Engineering, University of Minnesota, Minneapolis, Minnesota 55455, USA}
	
	\begin{abstract}
	        Maxwell lattices are characterized by an equal number of degrees of freedom and constraints. A subset of them, dubbed topological lattices, are capable of localizing stress and deformation on opposing edges, displaying a polarized mechanical response protected by the reciprocal-space topology of their band structure. In two dimensions, the opportunities for topological polarization have been largely restricted to the kagome and square lattice benchmark configurations, due to the non-triviality of generating arbitrary geometries that abide by Maxwell conditions. In this work, we introduce a generalized family of augmented topological lattices that display full in-plane topological polarization. We explore the robustness of such polarization upon selection of different augmentation criteria, with special emphasis on augmented configurations that display dichotomous behavior with respect to their primitive counterparts. We corroborate our results via intuitive table-top experiments conducted on a lattice prototype assembled from 3D-printed mechanical links.
		\vspace{0.4cm}
	\end{abstract}
	
	\maketitle

Mechanical metamaterials feature dynamical behavior that transcends that of conventional elastic media~\cite{babaee20133d, rafsanjani2016bistable, meeussen2020topological, zunker2021soft}. Lattice metamaterials, specifically, owe their distinctive macro-scale behavior to the periodicity of their microstructure, composed of tessellated \emph{unit cells} \cite{brillouin1953wave, phani2006wave, widstrand2022bandgap}. Maxwell systems are a special class of lattices, characterized by an equal number of degrees of freedom and constraints \cite{maxwell1864calculation, calladine1978buckminster, pellegrino1986matrix}. They posses interesting characteristics such as reconfigurability \cite{guest2003determinacy, sun2012surface, mao2013effective, rocklin2017transformable}, geometric behavioral dualities \cite{gonella2020symmetry, fruchart2020dualities}, and, of current relevance, the ability to host zero-energy (\emph{floppy}) modes \cite{pellegrino1993structural, lubensky2015phonons, mao2018maxwell} -- where sites displace without straining the bonds.

When freed from an infinite domain by cutting certain bonds, Maxwell lattices localize zero modes at the edges due to a local imbalance between degrees of freedom and constraints \cite{lubensky2015phonons, lubensky2015phonons,  mao2018maxwell}. Certain Maxwell lattices have been shown to localize deformation and stress on opposite edges, a feature referred to as \emph{topological polarization}, which is protected against defects by the momentum-space topology of the bulk \cite{hasan2010colloquium, qi2011topological, kane2014topological, chen2014nonlinear, paulose2015topological, xin2020topological, widstrand2022stress}. This polarization manifests as an excess of zero modes at the \emph{floppy edge}, while the opposite edge remains rigid. 

Topological polarization has been discussed for lattices with one-dimensional (1D) \cite{kane2014topological, chen2014nonlinear}, 2D \cite{kane2014topological, paulose2015selective, rocklin2016mechanical, rocklin2017transformable, ma2018edge, stenull2019signatures, pishvar2020soft, chapuis2022mechanical}, and 3D periodicity \cite{stenull2016topological, bilal2017intrinsically, baardink2018localizing}, and, more recently, for 2D-periodic bilayers embedded in 3D space \cite{charara2021topological, charara2022omnimodal}. However, in 2D, studies have largely been restricted to deformed configurations of the canonical square \cite{rocklin2016mechanical, chapuis2022mechanical} and kagome lattice \cite{lubensky2015phonons, rocklin2017transformable}. A gap exists in the study of topological lattices with increased cell geometry complexity and kinematics. Filling this gap provides a golden opportunity to search into a broader design landscape for additional topologically polarized configurations. 

In this letter, we introduce a family of topological lattices whose augmented unit cells contain a higher number of sites and bonds than the kagome or square lattices. In general, the task of generating augmented unit cells adhering to Maxwell conditions is non-trivial, as configurations resulting from the process may be over- or under-coordinated. A promising path consists of leveraging known unit cells amenable to topological polarization as elementary building blocks. In this regard, here we adopt a distorted kagome cell as a primitive geometry, and augment it through a series of mirror-folding operations. We document analytically an emergent topological polarization, and explore parametrically its robustness upon changes in geometry. Although we primarily focus on single-mirror augmentation, we introduce a generalization of this approach to multi-mirror augmentation strategies; while generally not guaranteeing full polarization, these extended strategies still yield asymmetric response scenarios with an unequal number of zero modes localized at opposing edges. We validate our results via experiments on a prototype assembled from 3D-printed mechanical links.

\begin{figure*}[t]
    \centering
    \includegraphics{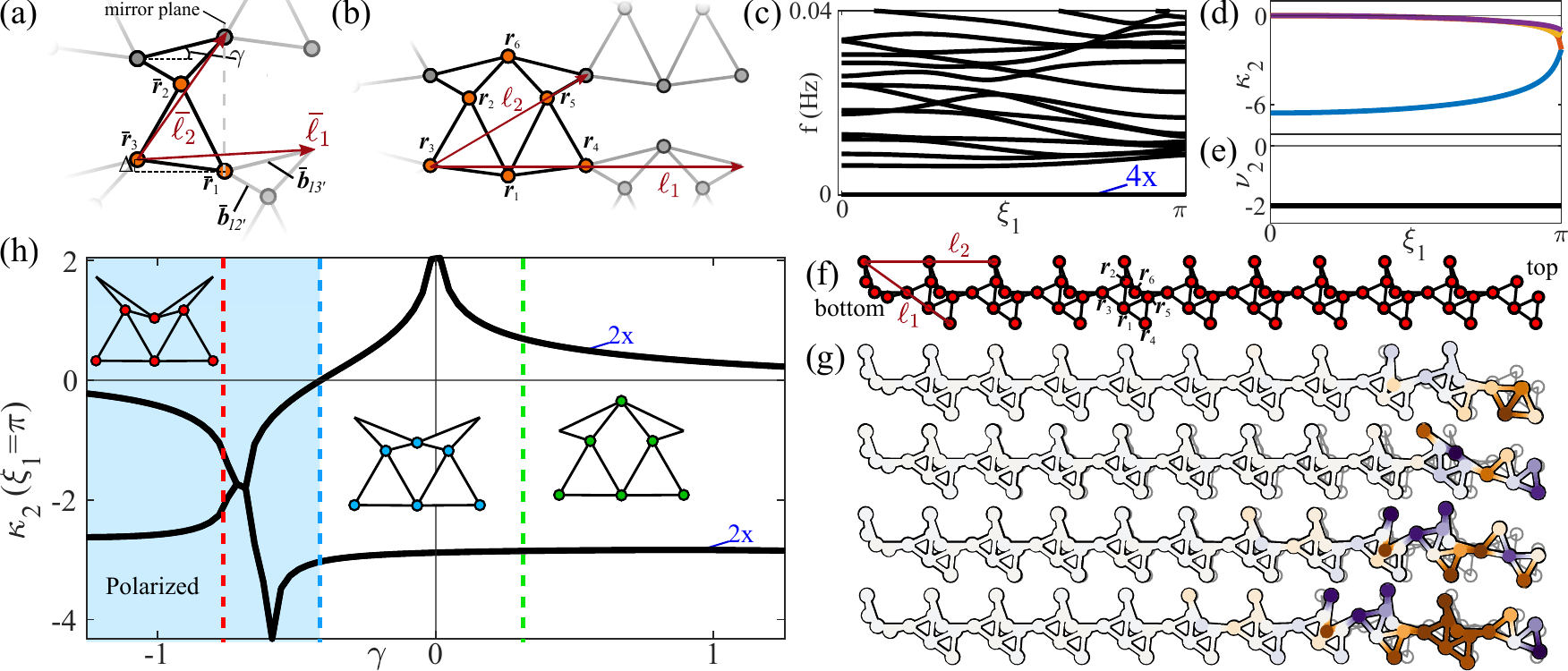}
    \caption{(a) Kagome (\SK{}) and (b) bi-kagome (\DK{}) unit cells, where orange sites belong to the unit cell and gray sites belong to adjacent unit cells; (c) supercell band diagram for the \DK{} unit cell with $\paramDeform=-\pi/4$, featuring four zero modes, and (d) decay rate $\kappa_2$ of zero modes and (e) winding number $\winding_2$ across the Brillouin Zone (BZ). (f) Supercell geometry and (g) mode shapes of the edge-localized zero modes at $\xi_1=\pi$. (h) $\kappa_2$ at $\xi_1=\pi$, as a function of $\paramDeform$, with examples of fully polarized (red), non-polarized (green), and phase transition (blue) configurations.}
    \label{fig:DKL}
\end{figure*}

A kagome unit cell consists of two triangles meeting at a vertex. Alternatively, it can be interpreted as three sites $\bar{\sites}_i$ connected to each other and to adjacent cells by six bonds $\bondKag_{ij}=\siteKag_j - \siteKag_i$, taken as springs of equal stiffness. A parametric description of the unit cell shown in Fig. \ref{fig:DKL}(a) is given in terms of $\siteKag_1=[1,\Delta]$, $\siteKag_2=[\cos{(\pi/3)},\sin{(\pi/3)}]$, and $\siteKag_3=[0,0]$, with lattice vectors $\bar{\latVecO} = [2,\sin{(\paramDeform)}-\Delta]$ and $\bar{\latVecT} = [1,\sin{(\pi/3)}+0.57\sin{(\pi/4 + \paramDeform)}-\Delta]$, where $\paramDeform$ is a parameter that rotates, counterclockwise, bonds $\bondKag_{13'}=(\siteKag_3 + \bar{\latVecO}) - \siteKag_1$ and $\bondKag_{12'}=(\siteKag_2 - \bar{\latVecT}+\bar{\latVecO}) - \siteKag_1$, changing the length of $\bondKag_{13'}$ (translating $\siteKag_3 + \latVecT$ vertically), and $\Delta$ adjusts the vertical position of $\siteKag_1$. Starting from this, we introduce the augmented cell shown in Fig. \ref{fig:DKL}(b), which we refer to as bi-kagome (\DK{}), acquired by mirroring the kagome cell about an axis (dashed line) passing through $\siteKag_1$ and $\siteKag_3+\bar{\latVecT}$. This geometry features six sites $\site_i$ and 12 bonds $\bond_{ij}=\site_j - \site_i$ per unit cell, with periodicity over $\latVecO=[4,0]$ and $\latVecT=[2,\sin{(\pi/3)}+0.57\sin{(\pi/4 + \paramDeform)}]$, where $\site_1=\siteKag_1$, $\site_2=\siteKag_2$, $\site_3=\siteKag_3$, $\site_4=[2,0]$, $\site_5=\site_2+[1,0]$, $\site_6=[1,\sin{(\pi/3)}+0.57\sin{(\pi/4-\paramDeform)}+\sin{(\paramDeform)}]$. For clarity, henceforth we refer to the primitive kagome as single kagome (\SK{}). \keepOrDiscard{Within the available landscape of augmentation and parametrization techniques, the current method is chosen as only one of the \DK{} lattice vectors is dependent on $\paramDeform$}. In this work, for convenience, we fix $\Delta=-0.03$, reducing the parameter space to just $\paramDeform$ (details about the effects of varying the parameter $\Delta$ can be found in the Supplemental Materials (SM) \cite{[{See Supplemental Materials at }][{ for details about: the effects of varying the parameter $\Delta$,  exploration of the topological polarization of the \DK{} lattice in the $\latVecO$, video of experiments, calculations of zero mode decay rates and winding number, $\paramRot$ \DK{} parametrization and information on its polarization, density calculation, effects of $T_h$ on band diagram, examples of $T_d$ diagonal mirroring geometries.}]supplemental}). 

We investigate the polarization of the augmented cell by adapting a general framework for Maxwell lattices \cite{mao2018maxwell}. \keepOrDiscard{The compatibility matrix $\compatibility$ of a cell relates site displacements $\Disp$ to bond elongations $\Elong$}. Maxwell lattices can experience zero-energy modes that displace the sites without deforming the bonds \keepOrDiscard{(i.e. $\compatibility\Disp=0$), therefore spanning the nullspace of $\compatibility$, under appropriate boundary conditions}. These zero modes can manifest in the bulk or localize to an edge. The topological polarization of a lattice is marked by an excess of zero modes localized on a certain edge. Focusing on the benchmark \DK{} configuration with $\paramDeform=-\pi/4$, we confirm the emergence of zero modes by studying the supercell shown in Fig. \ref{fig:DKL}(f), encompassing 10 cells along $\latVecT$, with Bloch-periodic boundary conditions along $\latVecO$ and free boundary conditions at the top and bottom edges. The resulting supercell band diagram (Fig. \ref{fig:DKL}(c)) features four overlapping zero modes. Their four corresponding mode shapes at $\xiO=\pi$, shown in Fig. \ref{fig:DKL}(g), indicate that \emph{all} four are edge modes localized on the top edge, denoting \emph{full polarization}. \updated{Note that although in the current work we explore polarization in the $\latVecT$ direction, the same treatment can be taken for $\latVecO$ (see SM for these details \cite{supplemental})}.

\begin{figure*}[t]
    \centering
    \includegraphics{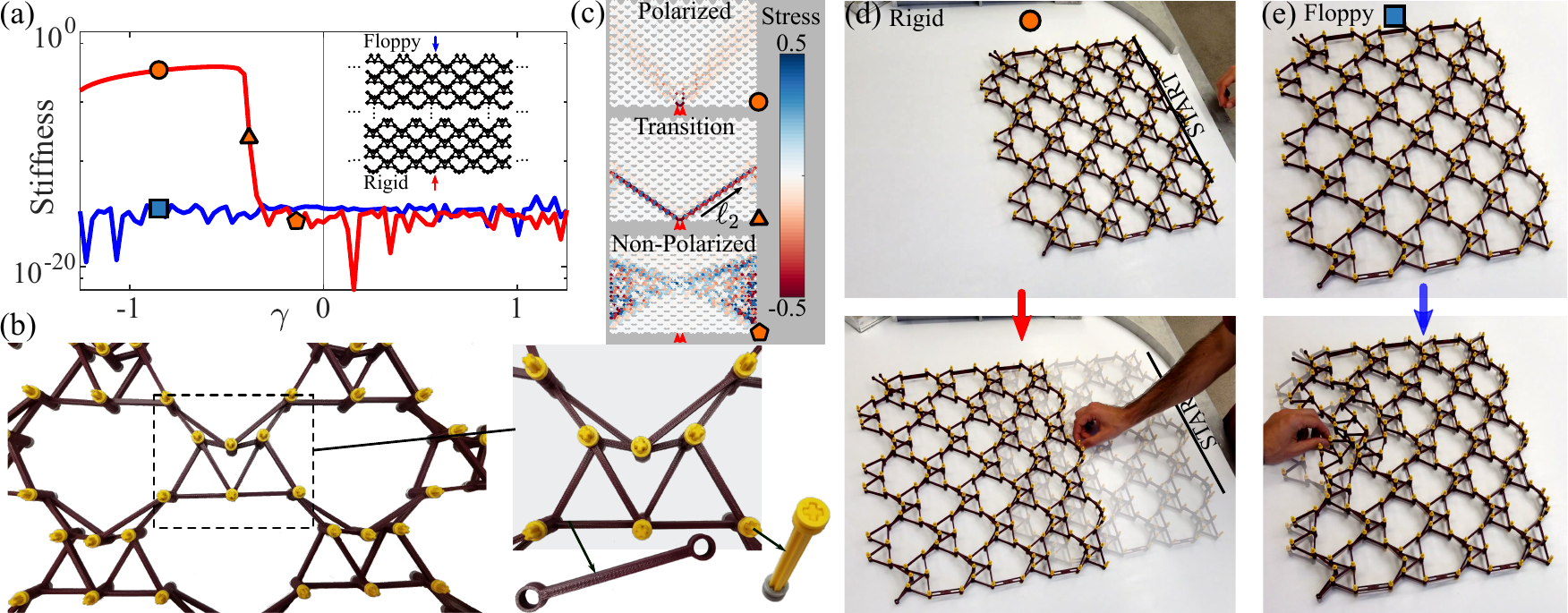}
    \caption{(a) Edge stiffness values, as a function of $\paramDeform$, inferred via full-scale static simulations of an $81\times81$ \DK{} lattice probed by a point load with the red (blue) curve referring to the rigid (floppy) edge. (b) Detail of the lattice with a close-up of an individual \DK{} unit cell, a 3D-printed link, and the Lego\textsuperscript\textregistered{} axles and bushings used as hinges. (c) Snapshots of stress concentration on an $11\times11$ lattice for a polarized, phase transition, and nonpolarized configurations with loading at the rigid edge. Quasi-static experiments performed loading from the rigid (d) and floppy (e) edges, yielding rigid-body motion and displacement localization, respectively (see video of experiments in SM \cite{supplemental}).}
    \label{fig:ES_Exp}
\end{figure*}

The degree of localization of a mode is characterized by its decay rate into the bulk $\kappa_2$ [along $\ell_2$], calculated for each transverse wave number $\xiO$, sampled along $\latVecO$, using the procedure described in Ref. \cite{charara2022omnimodal} (see SM for details on calculations of zero mode decay rates). Decay rates calculated across the BZ for $\paramDeform=-\pi/4$ are shown in Fig. \ref{fig:DKL}(d), where $\kappa_2<0$ ($>0$) denotes zero modes localized at the top (bottom) edge. The four zero modes are, indeed, localized on the top edge. Polarization along $\latVecT$ is protected by the topological invariant integer \emph{winding number} $\winding_2$, calculated for each wavenumber $\xiO$ over a closed loop in the BZ \cite{kane2014topological} (see SM for winding number calculation). As shown in Fig. \ref{fig:DKL}(e), $\winding_2=-2$ for all wavenumbers; this captures a transition [with respect to the ground state] of two zero modes from the rigid edge to the floppy edge, rendering the configuration fully polarized, and validates the topological roots of the observed behavior.

Focusing on the short-wavelength limit ($\xiO{}=\pi{}$), we study the dependence of the degree of \updated{localization of the modes (and therefore polarization of the lattice)} -- embodied by $\kappa_2$ -- on the parameter $\paramDeform$. The sweep $\paramDeform=[-\pi/2.5~\pi/2.5]$ is shown in Fig. \ref{fig:DKL}(h). For all values of $\paramDeform$, the decay rate curves are doubly degenerate (becoming quadruple degenerate locally around $\paramDeform=-\pi/6$), indicating that there exists, in general, two distinct decay patterns allowed by the lattice. Importantly, for configurations with $\paramDeform<-0.41$, the lattice displays full polarization, with all four zero modes localized on the top edge.

We quantify the asymmetry of the static response of the lattice by performing full-scale static simulations of an $81\times81$-cell domain, loaded by a unit-amplitude force applied at the midpoint of the top (bottom) edge while constraining the other edges. We infer edge stiffness by dividing the applied force by the resulting displacement on the loaded point. We repeat this exercise for the range $\paramDeform=[-\pi/2.5~\pi/2.5]$, with the stiffness of the rigid (floppy) edge shown in red (blue) in Fig. \ref{fig:ES_Exp}(a). For low $\paramDeform$ values, the large gap between the two curves denotes asymmetric edge stiffness: the floppy edge features essentially zero stiffness (within computational error), while the rigid edge value is finite. At $\paramDeform>-0.41$, the stiffness curves coalesce, marking the transition to a nonpolarized regime. The effects of a geometry sweep on the topological phase have been discussed in Ref. \cite{rocklin2017transformable} for a deformed kagome lattice, where the bounds of polarization were identified as the twist angles between the triangles for which their edges align to form states of self-stress (SSS). Here, the same condition cannot be met by the \DK{} lattice due to the impossibility to form straight lines of bonds for all $\paramDeform$. %
However, a qualitative rationale for the driver of phase transition can be inferred by studying the stress concentration patterns developing in the lattice in different regimes. Snapshots of static simulations of an $11\times11$ lattice loaded by a force applied at the rigid edge, shown in Fig. \ref{fig:ES_Exp}(c), reveal that, for a configuration at the phase transition point, \emph{force chains} develop from the loading point into the bulk along $\latVecT$, playing a role akin to an SSS. In stark contrast, a polarized configuration develops stress concentration at the rigid edge, decaying into the bulk, while for a nonpolarized configuration, stress concentration patterns emanate from the lateral boundaries.

We construct a mechanical prototype resembling the ideal spring-mass system used in our theoretical predictions, with slender links 3D-printed (Prusa MK3S) out of polylactic acid (PLA) and connected with nearly frictionless hinges of Lego\textsuperscript\textregistered{} axles and bushings. The assembled $5\times8$ lattice is shown in Fig. \ref{fig:ES_Exp}(b), where the nonperiodic bonds at the left and right boundaries help eliminate trivial floppy behavior of dangling bonds. Figs. \ref{fig:ES_Exp}(d) and (e) show the lattice behavior when quasi-statically manually loaded from the rigid and floppy edges, respectively. When pushed from the floppy edge, the lattice responds by developing a soft mode that localizes at the floppy edge, eventually morphing into geometrically non-linear deformation, involving macroscopic rotations of the triangles, that decays sharply into the bulk. Note that topological polarization theory, which is strictly linear, predicts only the softness of an edge and not the nonlinear deformation regime that unfolds. In contrast, loading from the rigid edge results in rigid body motion of the entire lattice.

\begin{figure}[t]
    \centering
    \includegraphics{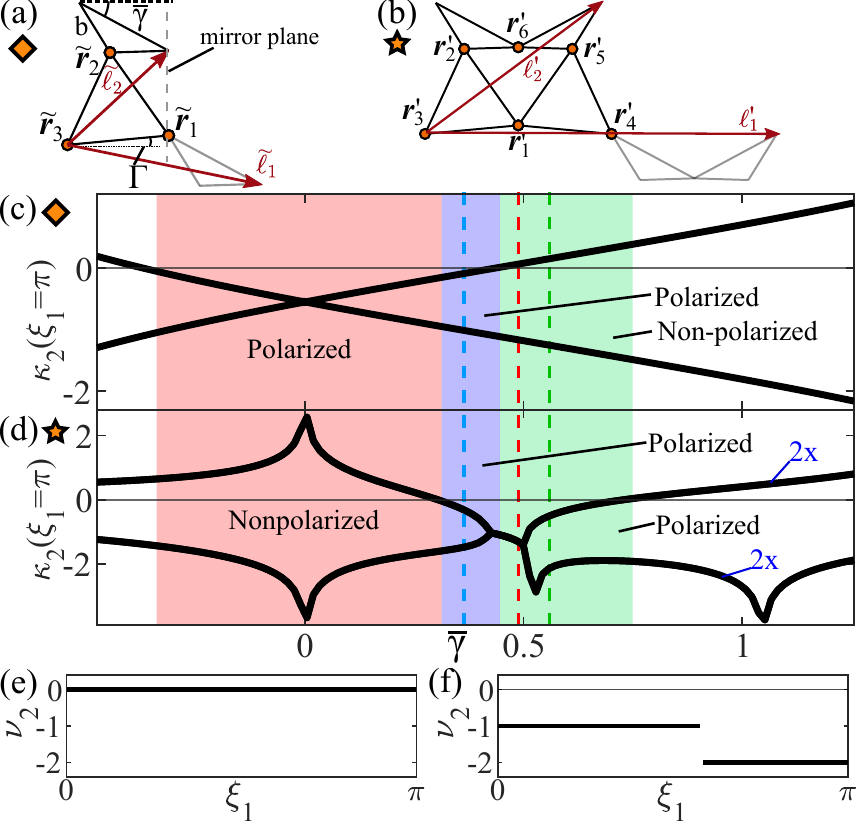}
    \caption{(a) \SK{} and (b) \DK{} unit cells parametrized in terms of the twist angle $\paramRot$. $\kappa_2(\xi_1=\pi)$ for the (c) \SK{} and (d) \DK{} in the interval $\paramRot=[-\pi/6.6~ \pi/2.5]$, where the red (green) shading highlights parameter regions where the \SK{} (\DK{}) is fully polarized while the \DK{} (\SK{}) is not, and the blue region highlights parameters where both geometries are polarized; winding number $\winding_2$ of the (e) \SK{} and (f) \DK{} lattices for an example case (green dotted line).}
    \label{fig:RotKag}
\end{figure}

To showcase the richness of strategies for cell augmentation available within the proposed framework, we consider an alternative variant of the \DK{} lattice, which, remarkably, can display polarization when its \SK{} counterpart is unpolarized. Take the \SK{} lattice and its \DK{} counterpart in Fig.s \ref{fig:RotKag}(a) and (b), respectively (see SM for details about $\paramRot$ \DK{} parametrization). Following the procedure above, we calculate the $\kappa_2$ for the \SK{} and \DK{} lattices over the parameter space $\paramRot=[-\pi/6.6~\pi/2.5]$, and plot the results for $\xiO=\pi$ in Fig. \ref{fig:RotKag}(c) and (d), respectively. The green region denotes a parameter range in which the \DK{} lattice is polarized while its \SK{} counterpart is not, and the red region exhibits the opposite trend, with the \SK{} polarized and the \DK{} nonpolarized. In the blue region, both lattices display polarization. As an example, winding number $\winding_2$ calculations for $\paramRot=0.57$ (green dotted line) for the \SK{} (Fig. \ref{fig:RotKag}(e)) and \DK{} (Fig. \ref{fig:RotKag}(f)) elucidate their topological dichotomy. For the \SK{} lattice $\winding_2=0$, indicating no polarization, while for the \DK{} $\winding_2=-2$ for shorter wavelengths and $\winding_2=-1$ for longer wavelengths, both signaling topological polarization, albeit with differing strengths. The integer change observed in $\winding_2$ is a well documented occurrence associated with the existence of a \emph{Weyl point} in the bulk band diagram, which denotes the existence of bulk zero modes \cite{rocklin2016mechanical, baardink2018localizing}. 

\begin{figure}[b]
    \centering
    \includegraphics{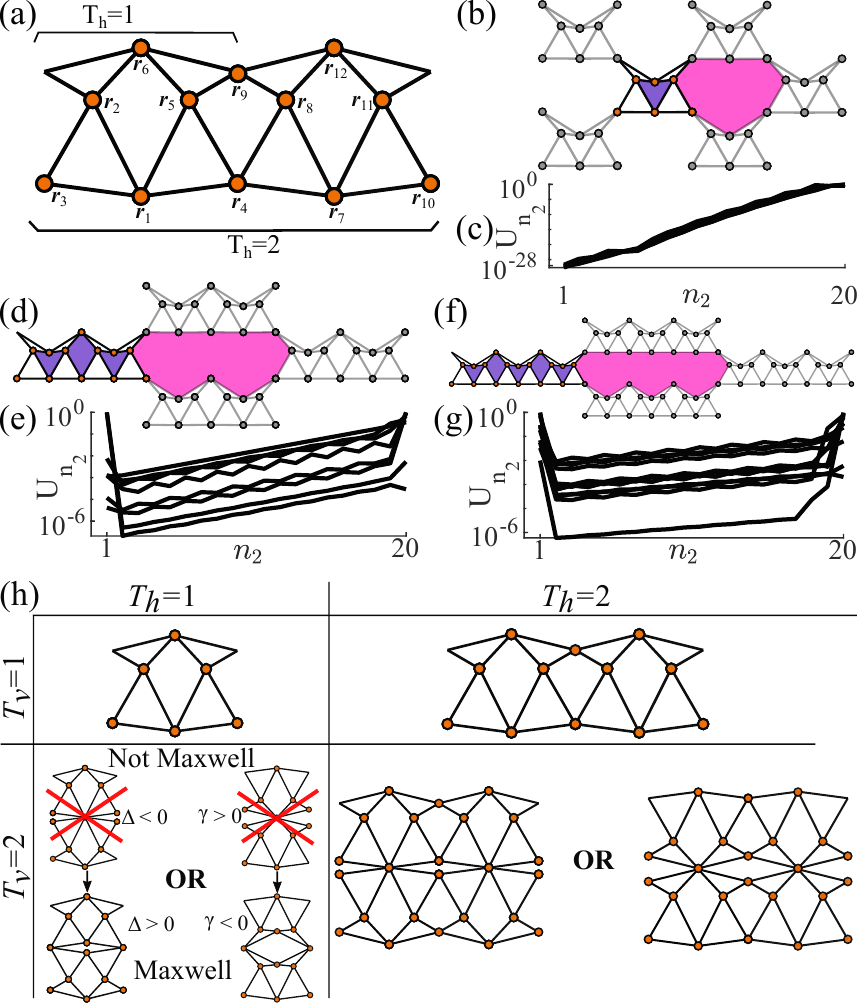}
    \caption{(a) Example of augmented unit cell featuring two connected \DK{} unit cells (i.e. $T_h=2$). Truncated lattices for (b) $T_h=1$, (d) $T_h=2$, and (f) $T_h=3$, highlighting emergent void polygons, and their modal displacement $U_{n_2}$ (c, e, g, respectively) plotted as a function of the unit cell index for all supercell zero modes. (h) Sample augmentation strategies with $T_h=1,2$ and $T_v=1,2$, where the $T_v=2$ row depicts augmentation via two distinct mirroring operations about the $\latVecO$ axis.}
    \label{fig:MTK}
\end{figure}

The versatility of the cell augmentation framework can be appreciated by considering a generalization of the process involving multiple mirroring steps preceding cell tessellation, resulting in additional proliferation of sites and bonds in each unit cell. For example, Fig. \ref{fig:MTK}(a) shows a unit cell generated by further mirroring the \DK{} unit cell from Fig. \ref{fig:DKL}(a) about sites $\site_4$ and $\site_3+\latVecT$ (the latter becoming $\site_9$ after mirroring). In principle, the process can involve an arbitrary number of horizontal mirroring operations; we introduce label $T_h$ denoting the number of \DK{} units assembled along the $\latVecO$ direction in the augmented cell. Fig.s \ref{fig:MTK}(b), (d), and (f) show three examples ranging from $T_h=1$ (original \DK{}) to $T_h=3$. The modal displacement $U_{n_2}$, defined as the magnitude of the site displacements in each cell of a given supercell modeshape, allows for qualitative visualization of the decay into the bulk of that mode. We plot $U_{n_2}$ for the considered configurations in Fig. \ref{fig:MTK}(c), (e), and (g), respectively, using a 20-cell supercell, where $n_2=1$ ($n_2=20$) denotes the bottom (top) edge. Interestingly, in the bulk, for $T_h>1$ the trend of localization towards the top edge persists, signaling robustness of the polarization against the augmentation process. However, most modes also display a sharp increase in displacement localized strictly to the last unit cell on the bottom edge. For augmented cells, this trivial edge effect may overwhelm the edge-selective attribute of the bulk that is ascribable to topological protection.

Further, it can be shown that, despite profound changes in geometry induced by varying $T_h$, the effective spatial density of the lattice remains constant (see SM for details of density calculation). This implies that we can sweep the structural design landscape of the lattice via cell augmentation, tuning the polarization, without changing the effective weight, which is often a major constraint in structural design. However, the augmentation process exacerbates the degree of heterogeneity of the lattice: as $T_h$ increases, the number of purple quadrilaterals follows $2T_h-1$ and the number of edges of the pink polygonal gaps is $4+4T_h$. This trend progressively localizes the solid phase of the lattice, while introducing (and progressively stretching) a length scale (i.e., $\ell_i$) associated with periodicity of the augmented cell (with interesting implications on bulk phonons -- see Supplemental Material for related band diagrams).

Finally, mirroring is not restricted to the $\latVecO$ direction; in fact, we can also introduce $T_v$ vertical mirrors (in the $\latVecT+\latVecO/2$ direction), or even $T_d$ diagonal mirrors (see SM for example geometry). Fig. \ref{fig:MTK}(h) tabulates unit cells generated for $T_h=1,2$ and $T_v=1,2$, where the row $T_v=2$ displays two alternative mirroring strategies. Frame ($T_h=1$~-~$T_v=2$) highlights that this augmentation strategy should be handled with care, as certain $\Delta$ and $\paramDeform$ ranges may result in non-Maxwell lattices. The same two vertical mirroring strategies are shown for ($T_h=2$~-~$T_v=2$), where the resulting lattices are still Maxwell by considering the \emph{average} count of sites and bonds over the cell, while accepting that certain sites will be under- and over- coordinated. Validating the topological nature of such resulting configurations requires future investigation.

In summary, we introduce a family of augmented Maxwell lattices obtained through mirror operations performed on kagome cells, which display full polarization. \updated{The mirror-folding strategy introduced is merely a small subset of a larger landscape of super-geometries that can be generated with the framework.} This process begins to uncover a larger dimensional space for topological lattices, with implications on the versatility of this geometry class for structural design.


\begin{acknowledgments}
This work is supported by the National Science Foundation (Grant No. EFRI-1741618). We acknowledge the support of the UMN Clifford I. and Nancy C. Anderson Student Innovation Labs and are grateful to James McInerney for the many insightful discussions.
\end{acknowledgments}

\bibliography{references.bib}

\end{document}


\title{Supplemental Material: A cell augmentation framework for topological lattices}
	\author{Mohammad Charara}
	\author{Stefano Gonella}
	\email{sgonella@umn.edu}
	\affiliation{Department of Civil, Environmental, and Geo- Engineering, University of Minnesota, Minneapolis, Minnesota 55455, USA}

	\maketitle

\section{Heatmap of multidimensional parameter space}\label{heatmap}

\begin{figure}[!htb]
    \centering
    \includegraphics{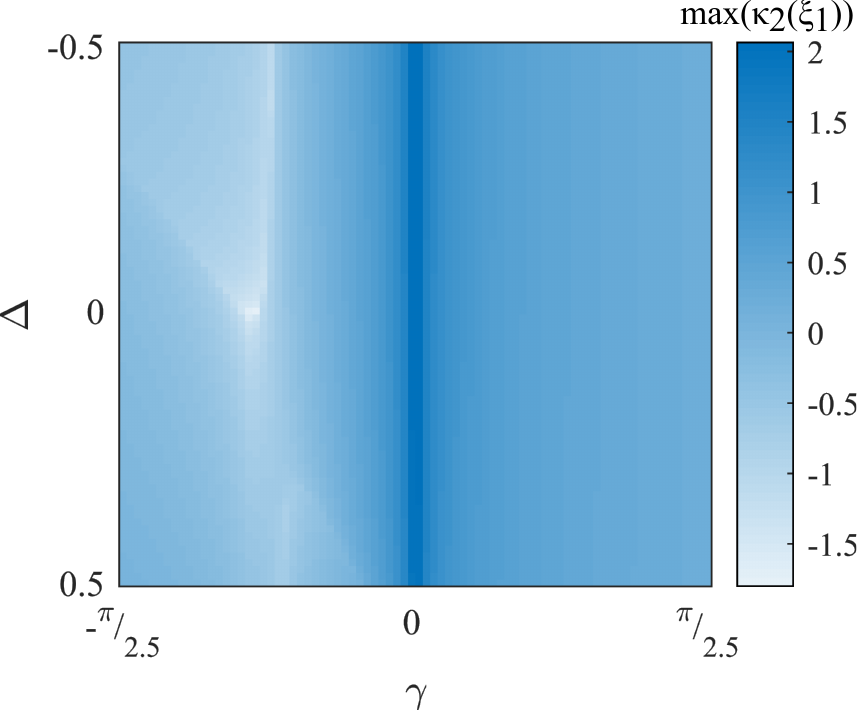}
    \caption{Heatmap of the decay rate $\kappa_2$ for the zero mode with lowest decay from the top edge at $\xi_1=\pi$ -- $max(\kappa_2(\xi_1=\pi))$ -- for all parameter combinations $\gamma=[-\pi/2/5~\pi/2.5]$ and $\Delta=[-0.5~0.5]$. Lighter color signifies a lower values (i.e. larger decay rate from the top and stronger polarization) while dark blue signifies a larger decay rate from the bottom edge and weaker/no polarization. }
    \label{fig:heatmap}
\end{figure}

Fig. \ref{fig:DKL}(a) and (b) in the main text introduce a parametrization of the \SK{} and \DK{} unit cells where $\gamma$ adjusts the angle of \SK{} bonds $\bondKag_{13'}=(\siteKag_3 + \bar{\latVecO}) - \siteKag_1$ and $\bondKag_{12'}=(\siteKag_2 - \bar{\latVecT}) - \siteKag_1$ counterclockwise (so that $\gamma=0$ results in an upper triangle that is isosceles) and $\Delta$ adjusts the height of \SK{} site $\siteKag_3$, with the corresponding changes mirroring for \DK{}. This parametrization is easily interpreted in the \DK{} as an effective vertical translation of $\site_6$. In the parameter ranges $\gamma=[-\pi/2.5~\pi/2.5]$ and $\Delta=[-0.5~0.5]$, we calculate the decay rates $\kappa_2$ of the zero modes of the \DK{} lattice at $\xiO=\pi$. We search for geometries where all four available zero modes are localized at the top edge; thus to confirm all $\kappa_2<0$, it is sufficient to check $max(\kappa_1(\xi_1=\pi))$: if this value is negative, then all modes are localized to the top edge -- we refer to  $max(\kappa_2(\xi_1=\pi))$ as the \emph{strength of polarization}. Results are summarized in the heatmap in Fig. \ref{fig:heatmap} where $max(\kappa_2(\xi_1=\pi))$ is documented for each parameter combination. The heatmap reveals that the geometry with highest polarization strength is where $\Delta$ gives the smallest perturbation, breaking the straight line of bonds along the pair of connected equilateral triangles - we remind the reader that in this work, we take $\Delta = 0.03$ for computations. Interestingly, if $\Delta=0$, the polynomial given by det$(\compatibility)$ reduces to an order lower than the number of available zero modes, therefore, in the analytical calculations $\Delta=0$ is avoided.

\section{Topological Analysis}
\begin{figure}[!htb]
    \centering
    \includegraphics{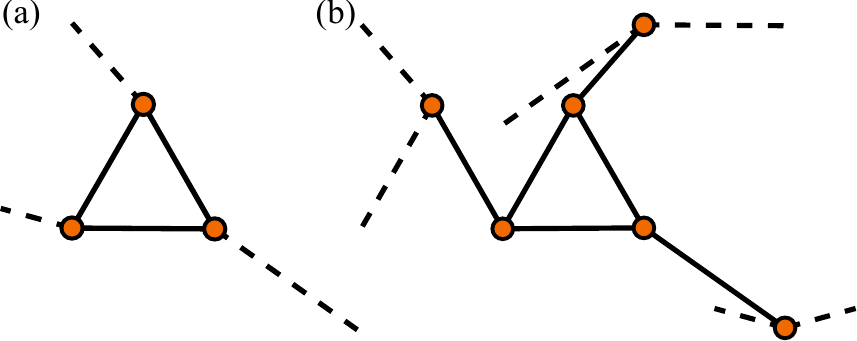}
    \caption{Example of (a) \SK{} and (b) \DK{} unit cell (with $\paramDeform=-\pi/4$) used in calculation of winding number $\winding_2$, where the solid lines are intracell bonds connecting sites in the unit cell to each other and dotted lines are intercell bonds that connect sites in the unit cell to sites in the adjacent unit cell. }
    \label{fig:symmUC}
\end{figure}

The compatibility matrix of a cell, $\compatibility$, relates site displacements $\Disp$ to bond elongations $\Elong$. The dynamical matrix $\dynamical=\compatibility^\dagger \stiff \compatibility$ -- where $\stiff$ is a diagonal matrix of bond stiffness constants (henceforth taken to be unity for convenience) -- yields eigenfrequencies $\omega$ and eigenvectors $\Disp$ that capture the mode shapes. The wave response of a periodic lattice is obtained from that of the unit cell upon applying Bloch periodic boundary conditions relating displacements at site $\site_i$ to those in site $\site_i+n_1\latVecO+n_2\latVecT$ as $\Disp(\mathbf{n})=\Disp e^{i\wavevec\cdot\mathbf{n}}$, where $\mathbf{n}=[n_1,n_2]$ identifies cell number $n_1$ ($n_2$) in the $\latVecO$ ($\latVecT$) direction and the components of the wavevector $\wavevec=(\xi_1,\xi_2)$ give~
the wave numbers defined along $\latVecO$ and $\latVecT$. The compatibility matrix $\compatibility(\wavevec)$ becomes Bloch-periodic and the eigenfrequencies $\omega$ and mode shapes $\Disp$ are the eigenvalues and eigenvectors of the dynamical matrix $\dynamical(\wavevec)= \compatibility(\wavevec)^\dagger\compatibility(\wavevec)$.

Zero modes in Maxwell lattices can present themselves in the bulk and/or localize at the edges. Recall, from the main text, that the topological polarization of a lattice is characterized by an excess of edge modes localized on one edge. To asses the degree of polarization of a lattice, we can compare the number of zero-modes that localize on (or decay from) opposing edges. The number of zero modes localized along the edges of a finite lattice in the $\ell_i$ direction depends on the number of bonds connecting a cell to its neighbor along that lattice vector -- four along the $\latVecT$ direction in the case of the \DK{} unit cell -- as confirmed by examining the nullspace of $\compatibility$. \updated{Generally, in an nonpolarized lattice -- more specifically in a nonpolarized lattice direction -- we expect an equal number of edge modes to localize on opposing edges; in the current study, then, the \DK{} lattice, in its unpolarized state, is expected to have two edge modes at each edge in the $\latVecT$ direction. }
   
Wave numbers, characterizing the spatial frequency of waves in each lattice direction, are generally complex $\xi=k_i + i \kappa_i$, with $k_i$ and $\kappa_i$ capturing the oscillatory and decaying components of the wave in the $i$ direction, respectively. The decay rates of these edge modes are calculated by finding the wave vectors at which the det$\compatibility$ vanishes (i.e. det$(\compatibility(\xiO,\xiT)=0$). This operation boils down to a root finding effort, as det$(\compatibility(\wavevec))$ is a polynomial whose order matches the number of available zero modes, and whose roots capture the decay rates. In the current work, we seek the decay rates of zero modes along the $\latVecT$ direction, thus we search for solutions $\xiT$ for which the equation holds for a given $\xiO$, across the Brilloin zone (BZ), with $\kappa_2=$Imag$(\xi_2)$ providing the decay. Decay rates are negative (positive) for edge modes localized to the top (bottom) edge. 

Topological polarization in a specific lattice direction $\ell_i$ is typically confirmed by studying the winding number of a lattice's unit cell, which can be interpreted as giving the integer count of modes transported from one edge to the other, rendering the former \emph{rigid} and the latter \emph{floppy}. This calculation is performed along a closed contour in the BZ (typically defined as $[-\pi~\pi]$ in the wavevector direction corresponding to the lattice direction of interest $\xi_i$) for each wavenumber in the transverse direction $\xi_j$. Recall that in this work, we are interested in assessing the polarization in the $\latVecT$ direction, thus for each wavenumber $\xiO$ we calculate

\begin{equation} \label{eq:winding}
    \winding_2(\xiO) = \frac{1}{2 \pi} \int_{-\pi}^{+\pi} d\xiT \frac{\partial}{\partial_{\xi}} \text{Im} \log \det \compatibility(\xiO,\xiT).
\end{equation}

\noindent In the spirit of the bulk-boundary correspondence, this calculation is performed over a unit cell (shown in Fig. \ref{fig:symmUC}) with periodic boundary conditions, yet it provides information about the behavior of the edges of a finite version of the lattice.

\section{Kagome Analysis for \DK{} comparison}

\begin{figure}[!htb]
    \centering
    \includegraphics{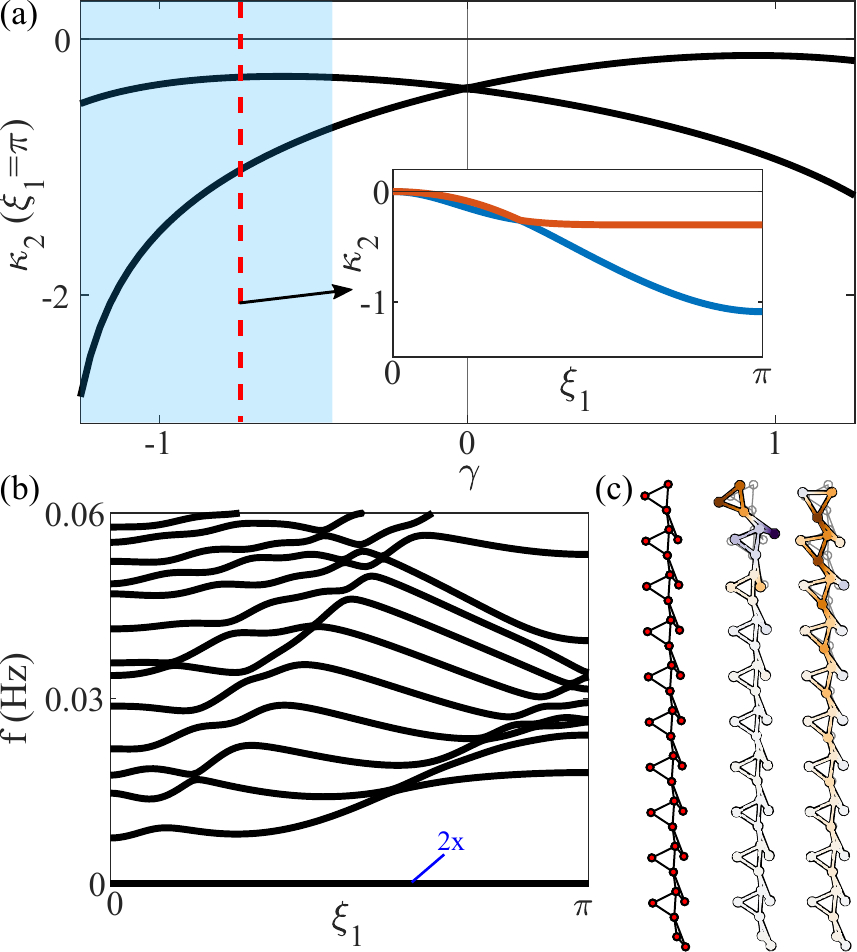}
    \caption{(a) Decay rates $\kappa_2(\xi1=\pi)$ for the kagome (\SK{}) lattice parametrized as shown in Fig. \ref{fig:DKL}(a) in the main text, for $\gamma=[-\pi/2.5,\pi/2.5]$, with the blue region in the plot serving as a reminder of the parametric region where the bi-kagome (\DK{}) unit cell displays full polarization towards the top edge. The inset shows decay rates of the zero modes across the entire BZ for the \DK{} parameter choice used in computations and experiments ($\gamma=\pi/4$, highlighted by the red vertical dotted line). (b) Band diagram for the \SK{} cell with $\gamma=-\pi/4$ with the (c) supercell used for its calculation (left) and corresponding mode shapes for the two available edge zero modes (middle and right).}    \label{fig:DeformSKL}
\end{figure}

To put the \DK{} lattice results from the main text in context, we find the decay rates $\kappa_2$ of a conventional kagome (\SK{}) unit cell for $\gamma=[-\pi/2.5~\pi/2.5]$, using the parametrization shown in Fig. \ref{fig:DKL}(a) in the main text. The decay rates of each geometry at $\xi_1=\pi$ (i.e. $\kappa_2(\xi_1=\pi)$) are shown in Fig. \ref{fig:DeformSKL}(a), where we see that, for all $\gamma$, the \SK{} is fully polarized. The inset shows the decay rates across the entire Brilloin Zone (BZ) for $\paramDeform=-\pi/4$. Bloch analysis over a 10 cell supercell gives the band diagram in Fig. \ref{fig:DeformSKL}(b), with two overlapping zero modes, whose mode shapes at $\xi_1=\pi$ are shown in Fig. \ref{fig:DeformSKL}(c), with the supercell shown on the left. The mode shapes show localization towards the top edge of the lattice.

\section{\DK{} lattice polarization in the $\latVecO$ direction}

\begin{figure*}[!htb]
    \centering
    \includegraphics{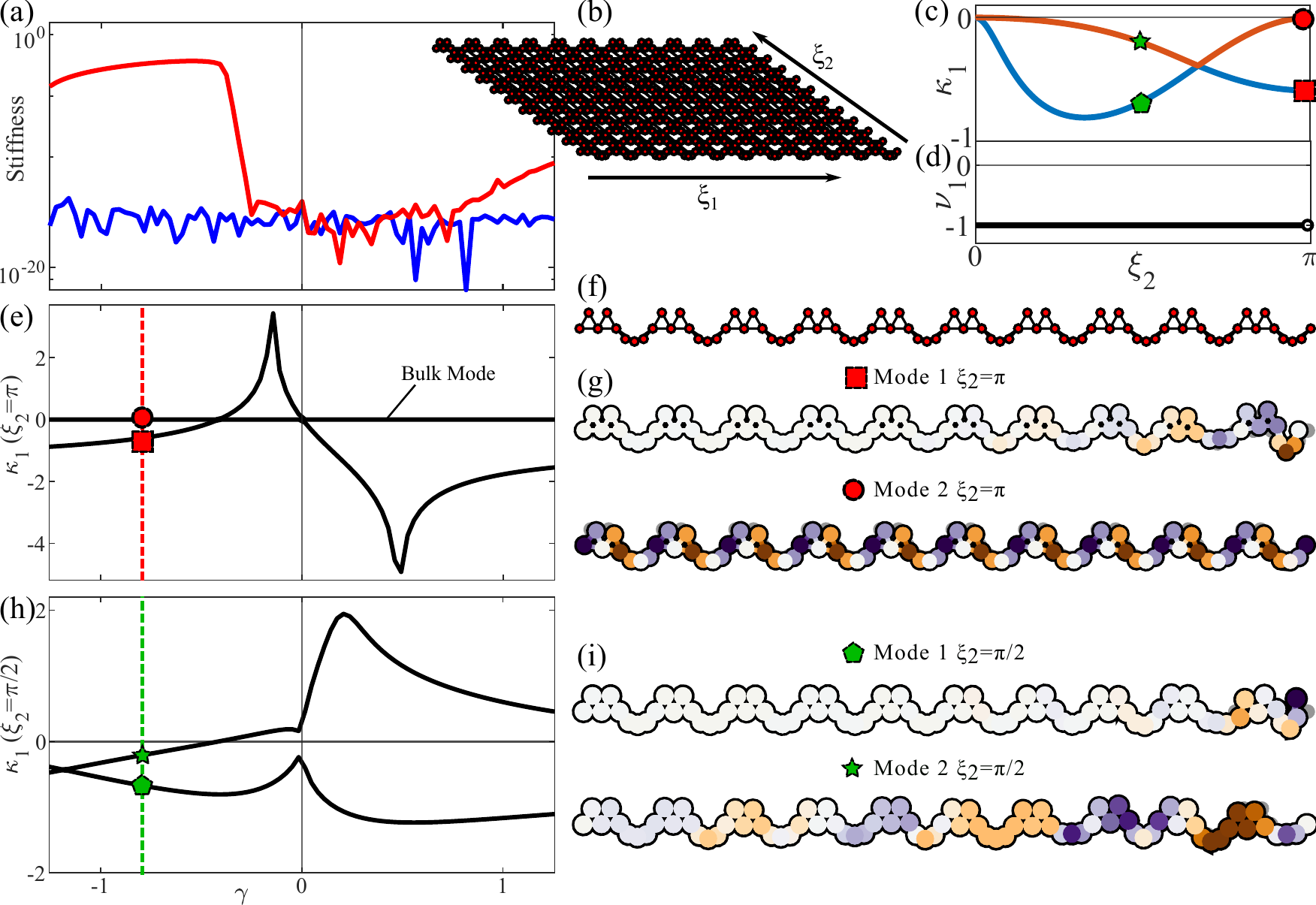}
    \caption{\updated{(a) Edge stiffness values, as a function of $\paramDeform$, inferred via full-scale static simulations of an $81\times81$ \DK{} lattice, with left and right edges parallel to $\latVecT$, probed by a point load with the red (blue) curve referring to the left (right) edge. (b) Sample \DK{} lattice (taken at $\paramDeform=-\pi/4$) constructed such that the left/right edges are compatible with $\latVecT$. (c) Decay rates $\kappa_1$ and (d) winding number $\winding_1$ for the \DK{} lattice with $\paramDeform=-\pi/4$. $\kappa_2$ at (e) $\xi_1=\pi$ and (h) $\xi_1=\pi$ , as a function of $\paramDeform$. (f) supercell constructed with periodic boundary conditions BCs) along $\latVecT$ (imposing $\xiT$) and open BCs along $\latVecO$ edges. Modeshapes of the zero edge modes for the \DK{} lattice with $\paramDeform=-\pi/4$ at (g) $\xiT=\pi$ and (i) $\xiT=\pi/2$, where the former case shows one localized and one bulk mode, and the latter show two localized modes.} }
    \label{fig:DKL_xiO}
\end{figure*}

\updated{An interesting aside is to evaluate the topological nature of the $\latVecO$ direction in the \DK{} lattice introduced in Fig. \ref{fig:DKL}. In this scenario the extreme edges in the $\latVecT$ feature open boundary conditions, an important detail when establishing edge modes. In the current work, we are interested in studying edge modes in a finite lattice; we identified the $\latVecT$ to be of interest because it features topological polarization, with all the zero edge modes from the bottom [rigid] edge transported to the top [floppy] edge. Thus, having  supercell set up in this fashion would allow visualization of edge modes in that direction.}   

\updated{In Fig. \ref{fig:DKL_xiO}(a) we plot edge stiffness comparisons between the right (blue) and left (red) edges across the $\paramDeform$ parameter space, with a sample \DK{} lattice at $\paramDeform=-\pi/4$ shown in Fig. \ref{fig:DKL_xiO}(b). Simulations reveal that the left edge displays a higher stiffness than the right at low $\paramDeform$ (shown as a gap between the blue and red curves). The dichotomy is therefore similar to that observed between the top/bottom edges; unlike the case in the main text, however, we note that the edge stiffness diverge slightly at larger values of $\paramDeform$.}
    
\updated{For the benchmark \DK{} with $\paramDeform=-\pi/4$, we perform further unit cell analysis showing decay rates $\kappa_1$ (Fig. \ref{fig:DKL_xiO}(c)) and winding number $\winding_1$ (Fig. \ref{fig:DKL_xiO}(d) across the Brillouin zone (BZ) for $\xiT$. Here, zero modes localized to the left (right) edge have $\kappa_1>0$ ($<0$) with a topological polarization to the left (right) signified by $\winding_1=1$ ($-1$). Note that here the nullspace of the compatibility matrix $\compatibility$ informs that two zero edge modes to localize along $\latVecO$, hence $\winding=\pm1$ in the polarized case. The current analysis confirms that both modes decay from the right edge ($\kappa_1<0$) for almost the entire BZ, except at $\xiT=\pi$. At the edge of the BZ, in this direction, there exists a Weyl point that takes the decay rate of one of the localized zero edge modes to 0. As a result, the winding number $\winding_1=-1$ for almost the entire BZ (confirming topological polarization to the right edge), but it is not well defined for $\xiT=\pi$, as shown by an open circle in Fig. \ref{fig:DKL_xiO}(d).}
        
\updated{Similar to the main text, we plot the $\kappa_1$ against $\paramDeform$ at $\xiT=\pi$ (Fig. \ref{fig:DKL_xiO}(e)). This shows that, across the $\paramDeform$ space, at $\xiT=\pi$, there exists one zero mode with no decay (i.e., bulk mode), and one zero mode localized to the right edge for almost the whole parameter space except for a small range below $\paramDeform<0$ (where it localizes to the left edge). We generate a supercell (Fig. \ref{fig:DKL_xiO}(f)) to calculate the mode shapes of these two zero modes for a sample geometry $\paramDeform=-\pi/4$ (shown in Fig. \ref{fig:DKL_xiO}(g)), which confirm that one of the modes localizes to the right edge while the other perfectly propagates through the bulk.}
    
\updated{We plot the $\kappa_1$ against $\paramDeform$ at $\xiT=\pi/2$ (Fig. \ref{fig:DKL_xiO}(h)) to inspect the zero edge modes' behavior away from the Weyl point at $\xiT=\pi$. Here, we see two distinct decay patterns, with both edge modes localizing to the right edge below $\paramDeform\approx-0.4$, with the remaining parameter space featuring one zero mode at each edge. We plot mode shapes of these two zero modes for a sample geometry $\paramDeform=-\pi/4$ (Fig. \ref{fig:DKL_xiO}(i)), confirming they are both localized to the right edge, albeit with one more heavily localized than the other.}

\section{Varying topological states in kagome and \DK{} lattices}

\begin{figure}[!htb]
    \centering
    \includegraphics{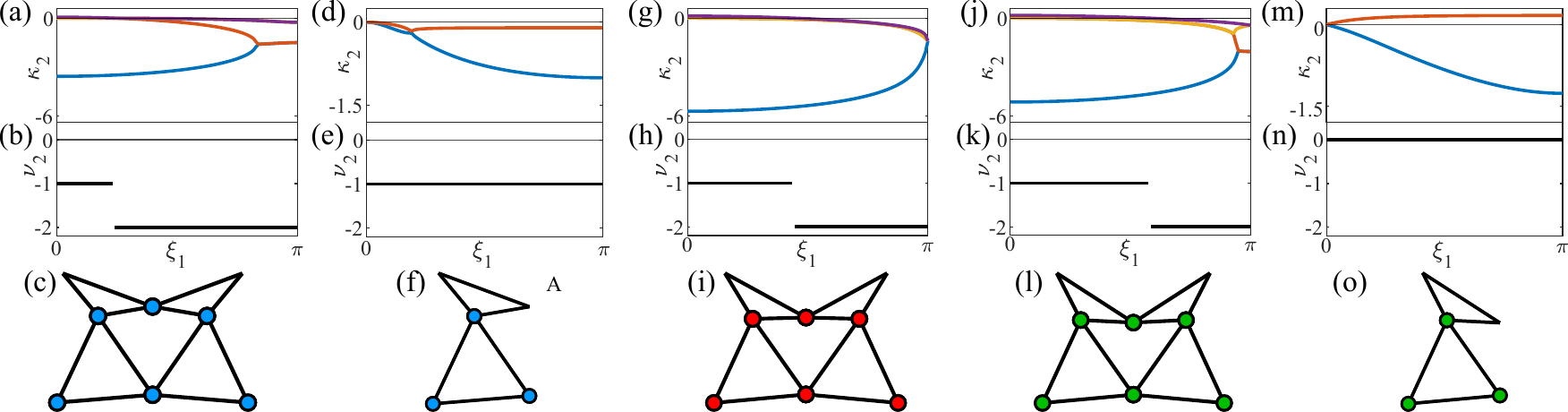}
    \caption{ (a, d) Decay rates $\kappa_2$ across the BZ, (b, e) winding number $\winding_2$, and unit cells for the (c) \DK{} and (f) \SK{} lattice, respectively, at the blue dotted line of Fig. \ref{fig:RotKag}(c); (g) $\kappa_2$ across the entire BZ, (h) $\winding_2$, and (i) unit cell for the \DK{} and \SK{} lattice in the red dotted line of Fig. \ref{fig:RotKag}(c), showing the geometry featuring the highest polarization strength; (j, m) $\kappa_2$ across the BZ, (k, n) $\winding_2$, and unit cells for the (l) \DK{} and (o) \SK{} lattice, respectively, at the green dotted line of Fig. \ref{fig:RotKag}(c).}
    \label{fig:RotKagIndDecay}
\end{figure}

In the main text, Fig.s \ref{fig:RotKag}(a) and (b) introduce an alternative parametrization of the \SK{} and \DK{}, where $\paramRot$ and $\Gamma$  enforce a net nonlinear rotation of the two triangles with respect to each other, where $\Gamma$ is a function of $\paramRot$ (i.e., $\Gamma(\paramRot)$ due to the mirror constraint about $\siteRotDK_1$ and $\siteRotDK_6$. The sites $\siteRot_i$ and lattice vectors $\latVecRot_j$ of the \SK{} in this parametrization are given by:

\begin{equation}
    \begin{aligned}
        &\siteRot_1=(\cos(\Gamma),\sin(\Gamma))\\
        &\siteRot_2=(\cos(\pi/3 + \Gamma),\sin(\pi/3 + \Gamma))\\
        &\siteRot_3=(0,0)\\
        &\latVecRot_1=(\cos(\Gamma)+\cos(\paramRot), \sin(\Gamma)-\sin(\paramRot))\\
        &\latVecRot_2 = (\cos(\pi/3+\Gamma)+b\cos(\pi/6-\paramRot),\sin(\pi/3+\Gamma)+b\sin(\pi/6-\paramRot))~.
    \end{aligned}
\end{equation}


\noindent where $\Gamma=\pi/3 + \arccos{(b\cos{(\pi/6-\paramRot)})}$. The sites $\siteRotDK_i$ and lattice vectors $\latVecRotDK_j$ of the corresponding \DK{} resulting from the mirroring are given by:

\begin{equation}
    \begin{aligned}
        &\siteRotDK_1=\siteRot_1\\
        &\siteRotDK_2=\siteRot_2\\
        &\siteRotDK_3=\siteRot_3\\
        &\siteRotDK_4=(2\cos(\Gamma),0)\\
        &\siteRotDK_5=(\cos(\Gamma)+\cos(\pi/3 - \Gamma),\sin(\pi/3 + \Gamma))\\
        &\siteRotDK_6=(\cos(\Gamma),\sin(\pi/3+\Gamma)+b\sin(\pi/6 - \paramRot)\\
        &\latVecRotDK_1=(2\cos(\Gamma)+2\cos(\paramRot),0)\\
        &\latVecRotDK_2 = (\cos(\paramRot)+\cos(\Gamma),b\sin(\pi/6 + \paramRot)+\sin(\pi/3+\Gamma))~,
    \end{aligned}
\end{equation}

\noindent where $b$ is the length of the shorter sides of the top isosceles triangle, and all other sides are of unit length.

Fig. \ref{fig:RotKag}(c) and (d) in the main text show decay rates $\kappa_2(\xiO=\pi)$ for the \SK{} and \DK{} lattices as a function of $\paramRot$, highlighting a blue and green region where both or only one of the two lattices is polarized, respectively, and specifically pointing out selected lattice configurations $\paramRot$ at the blue, red, and green dotted line. Here we show decay rates $\kappa_2$ and winding numbers $\winding_2$ for \DK{} (Fig.s \ref{fig:RotKagIndDecay}(a) and (b)) and \SK{} (Fig.s \ref{fig:RotKagIndDecay}(d) and E) configurations at the blue dotted line, whose unit cells are shown in Fig.s \ref{fig:RotKagIndDecay}(c) and (f), respectively. We see that, in the blue region, indeed, both geometries display polarization where the \DK{} has $\winding_2=-2$ ($\winding_2=-1$) at shorter (longer) wavelengths, while the \SK{} shows $\winding_2=-1$ across the entire BZ. Similarly, we show $\kappa_2$ and $\winding_2$, for \DK{} (Fig. \ref{fig:RotKagIndDecay}(j)) and \SK{} (Fig. \ref{fig:RotKagIndDecay}(m)) configurations at the green dotted line, with their unit cells shown in in Fig. \ref{fig:RotKagIndDecay}(l) and (o), respectively. Winding numbers $\winding_2$ for each configuration are reproduced from the main text in Fig.s \ref{fig:RotKagIndDecay}(k) and (n), respectively, for convenience. Here, decay rates corroborate the winding number discussion in the main text where the \DK{} lattice shows all four modes decaying from the same edge at shorter wavelengths, while the \SK{} shows one mode localized to each edge. Finally, for completion, we show the \DK{} geometry with highest strength of polarization -- defined as the configuration with the strongest zero mode decay signature from the top edge, mathematically seen as the lowest value of $max(\kappa_2(\xiO=\pi))$ --  (Fig. \ref{fig:RotKagIndDecay}(i)) and plot $\kappa_2$ and $\winding_2$ in Fig.s \ref{fig:RotKagIndDecay}(g) and (h), respectively -- in this local parametric region, small perturbations to $\paramRot$ have great impact on the polarization strength. It is interesting to note that all \DK{} geometries using this parametrization feature a Weyl point that progressively moves towards $\xiO=\pi$ as the parameter $\paramRot$ increases. Note that the zero mode decay rates trends remain unchanged for the \SK{} unit cell rotated $180^\circ$ about the $y$-axis.

\begin{figure*}[!htb]
    \centering
    \includegraphics{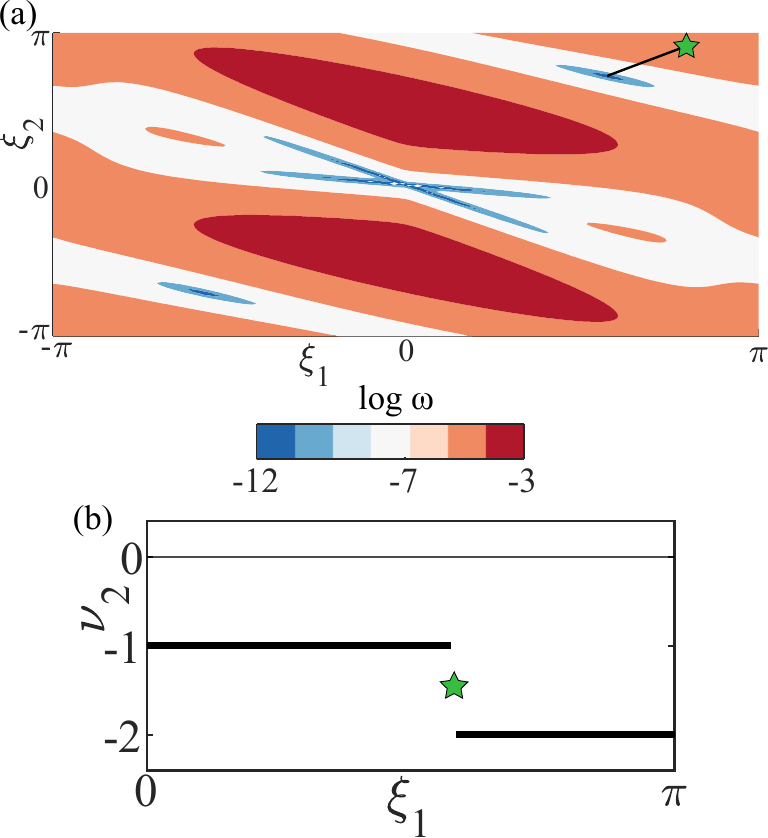}
    \caption{(a) Bulk band diagram for the \DK{} unit cell in Fig. \ref{fig:RotKagIndDecay}(l), where the Weyl point is highlighted by the green star; the effect of the Weyl point is seen on the (b) winding number $\winding_2$ for this unit cell, replotted here for convenience, highlighting the resulting integer change with a green star.}
    \label{fig:bulkWeyl}
\end{figure*}

\updated{To further display the effects of a Weyl point on the polarization of a lattice across the BZ, in Fig. \ref{fig:bulkWeyl}(a) we plot the first mode of the bulk band for the \DK{} sample geometry shown in Fig. \ref{fig:RotKagIndDecay}(l), with its resulting $\winding_2$ plot, recreated here for convenience, shown in Fig. \ref{fig:bulkWeyl}(b). Here, we highlight the Weyl point with a green star. Its effect can be seen on $\winding_2$, which changes by an integer amount on either side of the Weyl point's $\xiO$ location, a phenomenon we highlight with a green star matching the one on the bulk band.}
 
\section{Static \DK{} edge stiffness dichotomy}

\begin{figure}[!htb]
    \centering
    \includegraphics{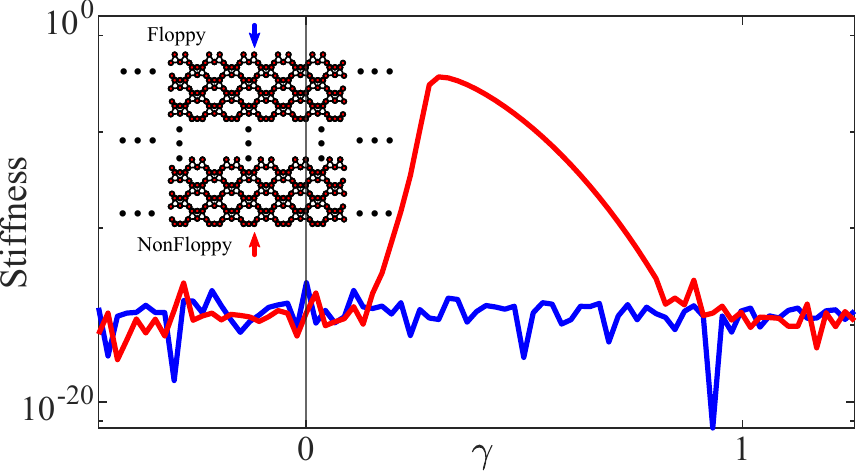}
    \caption{Stiffness comparison of the floppy (blue) and nonfloppy (red) edges, computed from static full-scale simulations, across the parameter space $\gamma=[-\pi/6.6~\pi/2.5]$, for the parametrization shown in Fig. \ref{fig:RotKag}(b) in the main text}.
    \label{fig:RotStatic}
\end{figure}

We perform computational static simulations to confirm the stiffness dichotomy between the floppy and rigid edges in a lattice composed of $21\times21$ \DK{} unit cells parametrized as shown in Fig. \ref{fig:RotKag}(b) in the main text. We apply a unit force at the top (bottom) edge on the centermost cell, and confine the displacement of the other edges. The stiffness of an edge is taken to be the unit force divided by the displacement of the site where the force was applied. The stiffness of the floppy (rigid) edge is shown in Fig. \ref{fig:RotStatic}(a) as a blue (red) curve for $\gamma=[-\pi/6.6,\pi/2.5]$. The rigid edge displays higher stiffness for the parametric region where the \DK{} lattice is fully polarized, and the stiffness of both edges is comparable elsewhere. 

For this parametrization of the \DK{} unit cell, the parametric region for which the two curves are gapped depends on the size of the lattice. Interestingly, as the lattice grows in size, the width of the region where the stiffness of the rigid edge is greater than the floppy edge decreases. This may be due to the presence of Weyl points in the band diagrams, where as its size grows, the lattice is able to support waves of longer wavelength, allowing for the appearance of the zero modes localized at the rigid edge to emerge. This does not necessarily invalidate the polarization of the lattices, as they still have $\winding_2<0$, but it means that the presence of a zero mode, regardless of wavelength, at the rigid edge overwhelms the edge-dichotomous response. Note that the same is not true for the \DK{} parametrization shown in Fig. \ref{fig:DKL}(b) in the main text, where the results remain qualitatively the same regardless of lattice size, and $\winding_2$ shows no Weyl points.

\section{Augmented unit cell design with 2D arbitrary periodicity direction}

\begin{figure}[!htb]
    \centering
    \includegraphics{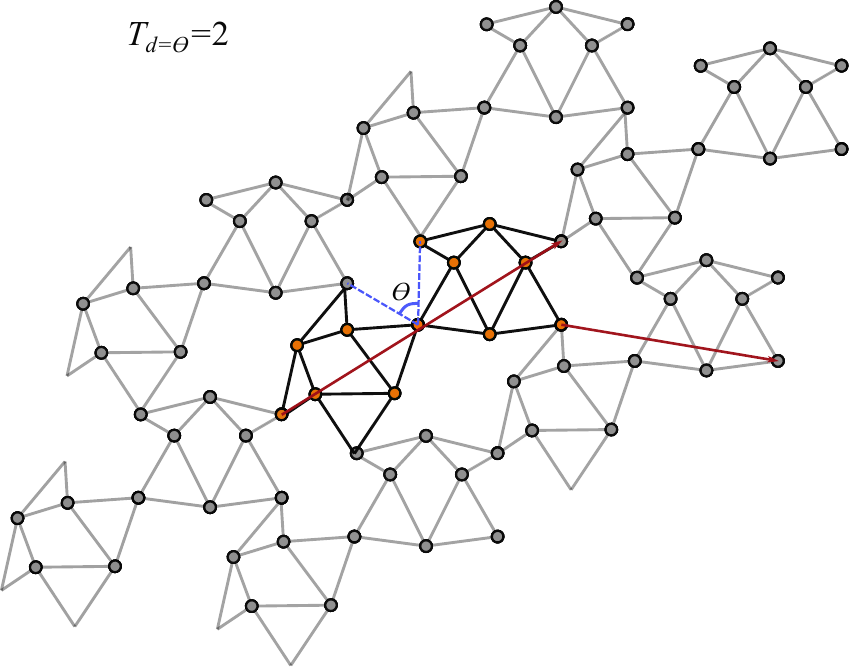}
    \caption{Augmented unit cell with arbitrary diagonal tessellation $T_d=2$ (i.e., two \DK{} units in the augmented cell) over an angle $\theta$, where maroon lines denote a choice of lattice vectors.}    \label{fig:ThTvTd}
\end{figure}

In the main text, we introduce a design scheme to generate augmented unit cells by tessellating in multiple directions, and discuss tessellation in the horizontal direction, parallel to $\latVecO$, with $T_h$ \DK{} units. Here, we briefly expand on vertical $T_v$ and diagonal $T_d$ mirrors. Fig. \ref{fig:MTK}(h) shows tabulated examples of $T_h$ ($T_v$) horizontal (vertical) tessellation of the \DK{} cell, and a combination of the two. Note that due to the restrictions of Maxwell criteria, in order to maintain an equal number of constraints and degree of freedom, tessellation cannot be performed in a trivial manner. Take, for example, tessellation ($T_h=1$~-~$T_v=2$). Tessellating the \DK{} cell in the top left corner of the table in Fig. \ref{fig:ThTvTd}(a) (with $\Delta<0$ on the left and $\paramDeform>0$ on the right), we cannot meet Maxwell criteria. This is true, for example, for all tessellations of odd valued $T_h$ when $T_v=2$. However, if we take $\Delta>0$ (for the mirroring scheme on the left) or $\paramDeform<0$ (for the mirroring scheme on the right), the tessellation becomes periodic. This highlights the need for careful parameter selection when using this framework to generate augmented unit cells adhering to Maxwell conditions.

In addition, tessellations need not be restricted to horizontal and vertical, and can in fact be performed in the diagonal direction to maintain some angle $\theta$ between the \DK{} sub-units, where the mirror plane becomes a linear combination of the two [original \DK{}] lattice vectors. The example in Fig. \ref{fig:ThTvTd} provides one version of what this may look like, where the unit cell is shown in black bonds and orange sites, the grayed out cells are adjacent, and the maroon arrows denote lattice vectors. Here, $T_d=2$, signifying that in this diagonally augmented cell, we have two \DK{} units.

We do not claim that all cells generated with this framework will give full polarization, or polarization at all. We simply introduce a method to generate augmented Maxwell unit cells. Further studies are required to understand the topological nature of this larger family of augmented lattices.

\section{Lattice Densities}
\begin{figure}[!htb]
    \centering
    \includegraphics{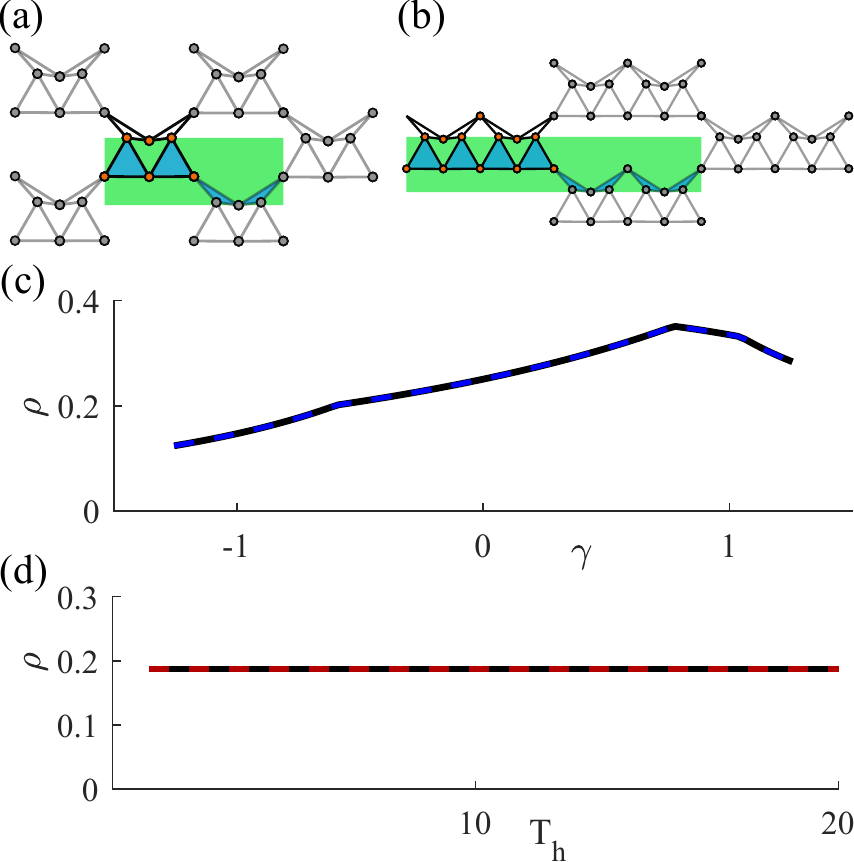}
    \caption{Examples of spatial areas used in calculating spatial density for the \DK{} lattice with (a) $T_h=1$ and (b) $T_h=2$, where the blue region is taken as the solid of a cell and the green box bounds the total area of a cell. (c) Spatial density of the \DK{} lattice, with the parametrization shown in Fig. \ref{fig:DKL}(b), as a function of $\paramDeform$ (solid black) with a comparison with its \SK{} counterpart (blue dotted line); (d) spatial density of the \DK{} lattice with $\paramDeform=-\pi/4$ as a function of $T_h$ (solid black) with a comparison of its \SK{} counterpart (red dotted line -- note that this line not plotted as a function of $T_h$ and only there for comparison). }    \label{fig:UCdensity}
\end{figure}

When using this framework in the design of lattices it is useful to understand how their spatial density changes upon transformation. Fig.s \ref{fig:UCdensity}(a) and (b) show a \DK{} lattice with $T_h=1$ and $T_h=2$, respectively. 
We consider the area taken up by the lattice $A_s$ as that encompassed by the triangles (blue shaded), and the total area footprint of the lattice $A_{tot}$ as maker by the green box bounded by the extremes in each cartesian dimension; the effective lattice density is then $\rho=A_s/A_{tot}$.  Fig. \ref{fig:UCdensity}(c) plots the density of the \DK{} lattice as a function of $\paramDeform$ (solid black line) and with a comparison to that of its \SK{} counterpart (blue dotted line) to show that they both maintain equal densities, which makes sense considering the fact that the \DK{} unit cell is mirrored from the \SK{}. Furthermore, in Fig. \ref{fig:UCdensity}(b) we check the density of the \DK{} lattice (solid black) as a function of $T_h$ for $\paramDeform=-\pi/4$ (results are exportable to a generalized value of $\paramDeform$) with a comparison of the \SK{} for control (red dotted) to show that, still, a constant lattice density is maintained as we expand the unit cell, despite the changes in shape and number of inter-cell polygonal gaps. Again, this makes sense as a simple mirroring multiplies, both, the area taken up by the lattice and the total area footprint. Note that using this logic, we can see that these results would carry over for arbitrary values of $T_v$, as well.

\section{Band diagrams for \DK{} lattices $T_h>1$}
\begin{figure}[!htb]
    \centering
    \includegraphics{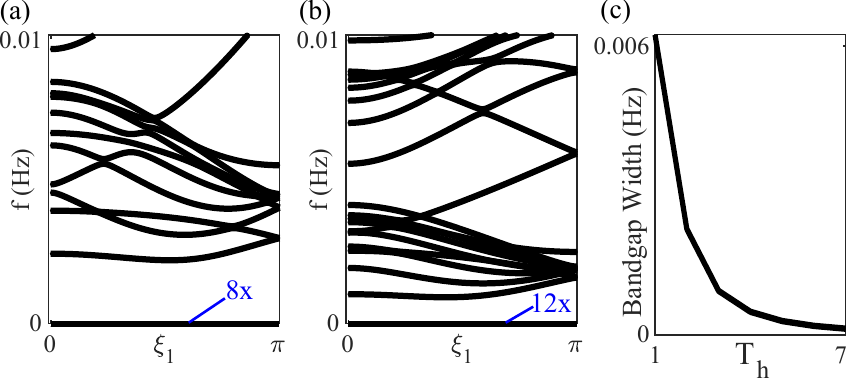}
    \caption{Band diagrams for augmented \DK{} lattices with (a) $T_h=2$ and (b) $T_h=3$. (c) Low frequency bandgap width in the augmented \DK{} lattice as a function of $T_h$. }    \label{fig:MTK_DBs}
\end{figure}

Band diagrams for the augmented \DK{} lattice with $T_h=2$ and $T_h=3$ are shown in Fig.s \ref{fig:MTK_DBs}(a) and (b), respectively. For reference, we remind the reader that the band diagram for the original \DK{} lattice (i.e., $T_h=1$) is shown in the main text in Fig. \ref{fig:DKL}(c). These band diagrams shown that, as $T_h$ increases, the low frequency band gap separating the zero modes and the bulk shrinks. This is no surprise considering the increase in size of the large polygonal void (shown in pink in Fig. \ref{fig:MTK}(b), (d), and (f) in the main text) with $T_h$, which suggests a decrease in the overall spectrum of the bulk (i.e., modes available at lower energies) due to the easier deformability of such lattices. An alternative way to look at this is to consider that the frequency range of the Bragg bandgaps in the augmented lattices -- arising from the periodic nature of their arrangement -- depends inversely on the length of the unit cell, whose length scale increases with $T_h$. Thus, as $T_h$ increases, the frequency range of such gap decreases. This lowering of the bulk band towards the zero modes is quantified by plotting the width of the low frequency band gap as a function of $T_h$ (Fig. \ref{fig:MTK_DBs}(c)), where we see the band gap shrink in size while the bulk approaches the zero modes.

\bibliography{references.bib}